\documentclass[a4paper, 11pt]{article} 

\usepackage[protrusion=true,expansion=true]{microtype} 
\usepackage[english]{babel}
\usepackage{graphicx} 
\usepackage{wrapfig} 
\usepackage{amsmath,amsthm,amssymb}
\usepackage{natbib}
\usepackage{mathpazo} 
\usepackage[utf8]{inputenc} 
\usepackage[footnotesize]{caption}
\usepackage{subcaption}
\usepackage{float}
\usepackage{booktabs} 
\usepackage{siunitx} 

\makeatletter
\renewcommand\@biblabel[1]{\textbf{#1.}} 
\renewcommand{\@listI}{\itemsep=0pt} 

\renewcommand{\maketitle}{ 
\begin{flushright} 
{\LARGE\@title} 

\vspace{50pt} 

{\large\@author} 
\\\@date 

\vspace{40pt} 
\end{flushright}
}

\newcommand{\p}{\mathbf{p}}
\newcommand{\D}{\mathbf{D}}
\newcommand{\C}{\mathbf{C}}
\newcommand{\V}{\mathbf{V}}
\newcommand{\A}{\mathbf{A}}
\newcommand{\btil}{\mathbf{\tilde{b}}}

\title{\textbf{Financial Contagion in Investment Funds}} 

\author{\textsc{Leonardo dos Santos Pinheiro} 
	\\{\textsc{Flávio Codeço Coelho}}
\\{\textit{Getulio Vargas Foundation}}} 

\date{\today} 

\begin{document}

\maketitle 



\begin{abstract}
Many new models for measuring financial contagion have been presented recently. While these models have not been specified for investment funds directly, there are many similarities that could be explored to extend the models. In this work we explore ideas developed about financial contagion to create a network of investment funds using both cross-holding of quotas and a bipartite network of funds and assets. Using data from the Brazilian asset management market we analyze not only the contagion pattern but also the  structure of this network and how this model can be used to assess the stability of the market.
\end{abstract}

\hspace*{3,6mm}\textit{Keywords:} financial networks , asset management , contagion 

\vspace{30pt} 


\section{Introduction}

In recent years the use of network representations for the study of economic systems has become ubiquitous, with many applications in the study of labor markets, micro-finance and production chains \citet{allen2008networks}. A special subject in these topics is the study of financial systems. Since the global financial crisis that hit the world in 2007-08 the interest in the intricate ways in which financial institutions are intertwined has soared, with studies showing the several facets of the interconnections of financial systems, specially in the way these connections affect global stability.

The 2007-08 crisis, which started in the US sub-prime mortgage market, rapidly spilled over to debt markets in a process of financial contagion that ultimately led to the demise of major American and European banks and triggered a world recession that spanned years. Interconnection is also a cause of major concern in the ongoing European debt crisis, with worries that the interconnection in the European bank system may cause a serious crisis if one nation defaults on its sovereign debt or enters into recession putting some of the external private debt at risk.

Due to the aforementioned events, much of the studies on the connectedness of financial institutions is focused on the mutual exposures between banks, specially the ones acquired on the interbank market (see \citet{cocco2009lending}, \citet{mistrulli2011assessing} and \citet{iori2006systemic}). But more recently some attention has been devoted to non-bank financial intermediaries, such as the studies being conducted by the Financial Stability Board to address what in being called the "Shadow Banking System" (\citet{board2011shadow} and \citet{board2011shadowstr}). In this work, we aim to explore one of these elements of financial systems, the asset management industry.

In the Brazilian market, in 2014,  asset management firms oversaw the allocation of approximately  U\$ 1T  in financial assets, consisting of a substantial part of the Brazilian financial system. Not only is the industry significant in size, but these firms and the funds they manage transact with other institutions in the financial system, and within themselves, in a variety of ways. As a consequence, this industry is heavily interconnected with the bank and insurance markets, augmenting greatly the effects it could potentially have if the industry were hit by a major financial shock.

While it is still highly debatable whether asset management in fact poses systemic risk, the industry has a number of factors that make it susceptible to financial shocks. Behaviors such as reaching for yield and herding, redemption risks associated to liquidity mismatch, high leverage and even behaviors of the asset managers can represent sources of risk (for a deeper discussion of the risks of asset management see the recent report of the \citet{ofr2013assets}). These are factors that have the potential to amplify financial shocks over the funds and, if the system is heavily interconnected, cause cascading failures and heavy losses to the entire financial system.

In this paper we develop a network model  to assess the inter-connectivity and how cascading failures can occur among investment funds. We also take a empirical approach to study the asset management market through simulations with data from the Brazilian Market. 

In a simple definition, networks can be described as collections of objects in which some objects can be connected forming a set of links. By this generic definition, many types of connections can be used to compose the edge's set (\citet{easley2010networks}).

The literature on financial networks contains many proposed metrics which can be used to define the connections between firms. For instance, \citet{huang2013cascading} propose the use of a bipartite network between firms and assets where a link represents a ownership relation between firm and asset, \citet{diebold2014network} propose connectedness measures built from pieces of variance decompositions and \citet{billio2012econometric} proposes a set of econometric measures based on principal-components analysis and Granger-causality networks. In this regards we follow closely the approach adopted by  \cite{elliott2014financial}, in which cross-holdings of organizations shares form the edges, taken together with the approach from \citet{huang2013cascading}, to better explain the effects of the contagion over common asset holdings as well.

By forming networks considering cross-holdings and common asset holdings we can analyze the structure of an investment fund network and look for potential impacts of financial shocks using contagion and diffusion models. While our model uses the ideas from the aforementioned works, this approach is not exhaustive, there are many other contagion and diffusion models in the finance literature  which could have been explored (see \citet{gai2010contagion} and \citet{allen2000financial}).

The structure of this paper is organized as follows. In Section 2 we provide a background and basic terminology on graphs and networks that will be used throughout the analysis. In Section 3 we present the contagion model using the frameworks developed  by \citet{elliott2014financial} and \citet{huang2013cascading}. Section 4 describes the data used and give a brief overview of asset management in Brazil. Section 5 presents the findings about the network structure and the results of simulations over the data. Finally, in Section 6 we discuss future research directions and present a summary and final remarks on the work.

\section{Graph Theory and Network Science}

In this section we provide some basic terminology about the concepts used along this paper. Its comprises concepts from graph theory for the representation of networks and from Social / Economic Network Analysis for understanding network structure and stability.

\subsection{Graph Theory}

A graph is a mathematical construct of a set of objects, called nodes or vertices, connected by a set of links, called edges. More formally a graph $G = (V, E)$ is an ordered pair consisting of a set of nodes $V$ and a set of edges $E$ where $E \subseteq V \times V $. The order of the graph is the number of nodes $n = |V|$ and its size is the number of vertices $m = |E|$.

A graph may be directed or undirected. A directed graph (or digraph) is a ordered pair $D = (V, E)$ consisting of a set of nodes $V$ and a set of edges $E$ where $E \subseteq V \times V $ and where, for every edge $(u, v) \in E$, there is a link that leaves \textit{u} and enters \textit{v}. We say that \textit{u} is the tail and \textit{v} is the head.

An import metric of networks is their degree distribution. The degree of a node $d_i$ is the number of connections it has. For a digraph we define in-degree as the number of connections incoming to a node and the out-degree as the number of connections leaving the node.

Graph labeling is the assignment of labels to edges and/or nodes of a graph. These labels often represent attributes of the graph. In a labeled graph we define the set of edges $E \subseteq V \times L \times V$ where $L$ is the set of labels. A \textit{weighted} graph is a labeled graph where edge labels are members of an ordered set, usually of integers or real numbers, which represent the "strength" of the connections. We define $w_{ij}$ as the weigh between nodes $i$ and $j$.

There are many ways to represent graphs. In this work, they are usually represented by their adjacency matrix. An adjacency matrix $M$, is a matrix whose entry $M_{ij} = w_{ij}$. In an unweighed graph we use $w_{ij} = 1$ by definition.

\begin{figure}
	\centering
    \includegraphics[width=0.5\textwidth]{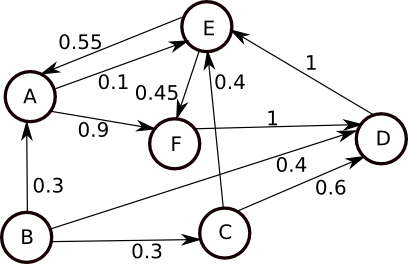}
    \caption{A weighted directed graph}
    \label{fig:weighted digraph}
\end{figure}

\subsection{Centrality Measures}

Graphs are a natural way to represent networks, but social economic networks usually exhibit some properties better analyzed through specific metrics. As was found by \citet{albert2002statistical}, many real networks exhibit a property  that their degree distributions  are scale-free. In social and economic networks, this property can create a structure where a few members of the network can gather most of the connections, thus, controlling the flow of information in the network.  This brings up the importance of analyzing the influence of members in a network. One way to analyze this importance is through centrality measures.

Centrality measures aim to describe how a given node relate to the network in some aspect of its structure, such as node position in the network. Four types of centrality measures are usually described in the network literature (\citet{jackson2008social}), each one aiming to describe an different aspect of node importance in the network. These are:

\begin{enumerate}
	\item Degree Centrality: How many connections a node has;
    \item Closeness Centrality: How easily a node can reach other nodes;
    \item Betweenness Centrality: How central the node is in creating paths between other nodes;
    \item Eigenvector Centrality: How important (well connected) the node's neighbors are.
\end{enumerate}

\textbf{Degree Centrality} is considered the most classical measure of centrality and  it measures how import a node is by the number of connections it has (\citet{freeman1980centrality}). The degree centrality is simply measured by $C_{d}(u) = d_i(u)$. While it is a relevant measure, it is considered limited for analyzing node influence. It's main flaw is that it misses the location of the node in the network while for some applications, like spread of processes, the position of the node in the network is a fundamental aspect.

\textbf{Closeness Centrality} measures a node importance by how close it is to any other node in the network (\citet{freeman1980centrality}) . One way to measure closeness is:

\begin{align}
	C_{c}(u) = \dfrac{n - 1}{\sum_{v=1}^{n-1} d(v,u)},
\end{align}

where $d(v,u)$ is is the shortest-path distance between $v$ and $u$, and $n$ is the number of nodes in the graph. In a diffusion process, the nodes with highest closeness centrality are likely to be affect by the process more rapidly than others.

\textbf{Betweenness Centrality} captures how well situated a node is in terms of the paths (\citet{freeman1980centrality}) in the network. The betweenness centrality of a node $u$ is the sum of the fraction of all-pairs shortest paths that pass through it. Mathematically we measure betweenness as::

\begin{align}
	C_{b}(u) = \sum_{s, t \in V} \dfrac{\sigma (s, t | u)}{\sigma (s, t)},
\end{align}

where $\sigma (s, t | u)$ is the number of shortest $(s, t)$-paths and $\sigma (s, t)$ is the number of those paths passing through $u$. In a diffusion process, a node that has high betweenness can control the flow of information in the network.

\textbf{Eigenvector Centrality} is based on the premise that a node's importance is measured by how important it's connections are (\citet{bonacich1972factoring}). The eigenvector centrality for node $u$ is $\mathbf{x}_i$ where $i$ is the index of node $u$ and $\mathbf{x}_i$ is the principal eigenvector in:

\begin{align}
	\mathbf{Ax} = \lambda \mathbf{ x} 
\end{align}

where $\mathbf{A}$ is the adjacency matrix of the network. In diffusion process, an node who is high on eigenvector centrality is connected to many nodes which themselves are connected to many nodes, thus multiplying their probability of contagion.

\subsection{Homophily and Assortativity}

Social and Economic networks sometimes exhibit a property that entities are more prone to establish connections with similar entities. This property was named homophily by \citet{kandel1978homophily}. Homophily may play an important role in economic networks, since it can mean a network can be largely segregated (\citet{jackson2008social}).

One way to measure homophily among labeled nodes is by assortativity. Assortativity measures the similarity of connections in a network with relation to some attribute or label.  We define the assortativity coefficient as: 

\begin{align}
	r = \dfrac{\sum_{i} m_{ii} - \sum_{i}  a_{i} b_{i}}{1 - \sum_{i} \ a_{i} b_{i}},
\end{align}

where $m_{ij}$ is the fraction of edges in the network that connect a type $i$ node to a type $j$ and $m_{ij} = a_{i} b_{j}$. The matrix M is the joint probability distribution (mixing matrix) of the specified attribute \citet{newman2003mixing}. Assortativity may play an important whole in financial networks as segregated financial firms may form regions of closure for the spreading of losses.

\subsection{Dynamic Networks}

Some networks vary over time.  The field that studies how networks change over time is known as network dynamics or dynamic network analysis (DNA). The main aspects of DNA is the analysis of the statistical properties of time varying networks and the simulation of network changes over time. DNA is a broad field and we refer to the work of \citet{carley2003dynamic} for those interested in it. For the purpose of financial networks and contagion we focus on the  stability of financial networks over time.

One way to measure network stability over time is to check the Jaccard similarity of nodes and/or edges between pairs of successive time spans (\citet{masys2014networks}).  This is a indication of how much nodes and/or edges are formed or removed from the network between small time steps. The Jaccard similarity coefficient is a measure of similarity between sets. The coefficient of a set S and a set T is $|S \cap T| / |S \cup T|$, that is, the ratio of the size of the intersection of S and T to the size of their union (\citet{rajaraman2012mining}).

\section{The Cross-Holdings Contagion Model}

In determining interconnectivity between investment funds we follow closely the model developed by \cite{elliott2014financial} for financial firms in which cross-holdings of shares among organizations may lead to cascading failures. In the context of investment funds, we measure cross-holdings of fund's quotas. We differ from the model when common assets holdings are considered in the process and in this regard we use the framework developed by \citet{huang2013cascading}.

\subsection{Primitive Assets and Cross-Holdings}

In the original model developed by  \cite{elliott2014financial} the value of organizations are determined by factors of production ( represented by the assets the organizations own)and the shares of the other organizations which they hold. There are $n$ organizations making up a set $N = {1, 2, ...,n}$. There is also a group of assets in the economy that may be owned by firms and these assets compose another set $M = {1,2,..., m}$. For investment funds as organizations, the primitive assets are formed by assets not issued by other investment funds and legally permitted to be bought by funds, these may be shares of companies, corporate bonds, government bonds, derivatives, among others. Funds can also hold shares of other funds, which creates the cross-holdings among organizations. As the funds are the only organizations in our system we shall use the terms interchangeably from this point.

The base value of a fund is determined by the value of it's assets. The value of an asset $k$ is denoted by $p_k$ and we call $\p$ the vector containing the values of the assets in $M$. We also call $\D$ the matrix which entry $D_{ik}$ is the share of the value of asset $k$ held by organization $i$. Complementary to $\D$ we have the matrix $\C$  in which, for each$,i,j \in N, C_{ij} \geq 0$ is the fraction of organization $i$ owned by organization $j$. The matrix $\C$ can be seen as a network of direct links between the organizations. There is also the share   $\hat{C}_{ii}  := 1 - \sum_{i \in N} C_{ij}$ of organization i which is not owned by other organizations in the system. This forms the matrix $\hat{\C}$. 

To determine the fair value of organizations, \citet{elliott2014financial} used a framework developed by \citet{fedenia1994cross} and \citet{brioschi1989risk}. The value $v_i$ of organization $i$ is determined by the value of it's assets plus the value of it's applications on other organizations:

\begin{align}
		v_i = \sum_k D_{ik} p_k + \sum_j C_{ij} v_j
	\end{align}

This equation can be expressed in matrix form by $\V = \D \p + \C \V$ and is solved to yield:

	\begin{align}\label{eq: cr1}
		\V = (\textbf{I} - \C)^{-1} \D \p
	\end{align} 

\citet{brioschi1989risk} and \citet{fedenia1994cross} argue that the true value of an organization is better captured by what is held by \textit{outside} investors. This value is equal to $\dot{v_i} = \hat{C}_{ii} v_i$, leading to:

	\begin{align}\label{eq: cr2}
		\dot{\V} = \hat{\C}(\textbf{I} - \C)^{-1} \D \p = \A \D \p
	\end{align} 

 We also call $\A = \hat{\C}(\textbf{I} - \C)^{-1}$ the \textit{ dependency matrix}. It captures the true value of the cross-holdings of quotas in the market and allow us to measure the true value of the funds and how changes in one organization's value shall affect any other.

\subsection{Firm Susceptibility to Financial Shocks}

Organizations can lose value in discontinuous ways under certain situations. We call these losses failure costs. Failure costs are assumed if an organization falls bellow some value threshold, in which case we can interpret that it has transitioned from a financially stable situation to an unstable one (\citet{elliott2014financial}). So, if organization $i$ value's fall below some threshold $\bar{v}_i$, it incurs in failure costs $\beta_i (\p)$. 

In a more general setting there are many possible explanations for failure costs. The main assumption in investment funds is that under certain situations there main occur a run on the fund, forcing it to sell assets at inopportune times at a discount rate, leading to the discontinuous losses. There may be many situations causing this run on the fund, such as reputation risk generated by the asset manager, performance risk caused by risky strategies and leverage, or economic risks such as financial bubbles and idiosyncratic shocks.

When funds are directly connected by cross holdings the spread of a discontinuous loss is straightforward. Nevertheless, failure costs incurred by one organization may "affect" other organizations  even if they are not directly connected in the network. The main way in which it can happen is common asset holdings. Since many assets are held simultaneously by many firms, the way one firm incurs in failure costs can force it to fire sale it's assets and force the price of the asset down in a way that other organizations holding the same assets may also face difficulties.

To correctly address the effect of common asset holdings we use the framework developed by \citet{huang2013cascading}. In addition to the network of cross-holdings described in Equation \ref{eq: cr1} we use a ancillary network formed by organizations as one type of node and assets as the other.

In this network a link between a fund and an asset exists if
the fund has the asset on its portfolio. It is a bipartite weighted graph where the weights on the links represent the gross value of the portfolio position in the asset.

Let $\mathbf{B}$ be the bipartite network between the organizations in \textbf{N} and the assets in \textbf{M}. We have that \[\mathbf{B} = \begin{bmatrix} 0_{n, n} & \mathbf{W} \\ \mathbf{W}^{\textnormal{T}} & 0_{m, m}      \end{bmatrix},\] where $\mathbf{W}$ is a $n \times m$ sub-matrix where $w_{ij}$ is the value of the position of fund $i$ in asset $j$. We can rewrite the term $\D \p$ in \ref{eq: cr2} in terms of $\mathbf{W}$. Let $\vec{1} = [1, 1, 1,..., 1]_{n}^{\textnormal{T}}$. We write: \[\D \p = \mathbf{W} \vec{1}. \] And Eq. \ref{eq: cr1} leading to:

\begin{align} \label{eq:cr3}
	\V = \A \mathbf{W} \vec{1}
\end{align}

In the presence of financial instability for an organization $i$, not only failure costs $\beta_i$ are incurred but also every asset $j$ owned by $i$ suffer a pressure to go down, becoming: $p_j = \dfrac{\sum_i w_{ij} - \omega w_{ik} }{\sum_i w_{ij}}$.

\subsection{The Contagion Model}

Financial contagion can occur when one organization fails and it's losses spread to other organizations causing them to fail as well. This can have the potential to generate a cascade of failures, potentially breaking the financial system as a whole. 

The financial contagion process can be described as a diffusion process where losses spread in the network of interconnections. Through both the cross-holdings connections and common asset holdings funds can be affected by this diffusion process.

If funds are directly connected, discontinuities will propagate in the cross-holdings network, affecting the final value of other funds invested in the broken firms in the path of connections. These affected funds, in turn, may start to face difficulties and cause new failures and discontinuous losses. At the same time, these funds may be forced to sell their assets, causing drops in asset prices which may also cause new failures and discontinuous losses. These losses shall propagate until a new equilibrium is reached.

But prior to contagion we must have a shock over the system. This shock can be of any kind, but for the contagion to be triggered at least one  fund must lose value until it fall bellow the critical value to move the system from equilibrium.

Bellow we describe an algorithm containing the step-by-step process to simulate the cascading failure process.

\begin{enumerate}
	\item Let $Z_t$ be the set of failed organizations at step $t$. We initialize $Z_0 = \emptyset$, indicating a starting state of equilibrium.
    
    \item At the initial moment we shock a market asset $m_i$ or a set of assets $\mathbb{M} = \{ m_1, m_2,..., m_k\}$ to be affected by the first shock (Affected assets depend on the nature of the shock). Each asset is reduced to a fraction of its original value $\eta p_i$, where $\eta < 1$ is determined by the strength of the shock.
    
    \item After the initial shock, the loss spreads in the network of cross-holdings and we recalculate the value of each fund in the system using  Equation\ref{eq:cr3},  checking which of them have fallen bellow a critical value $v_{\textnormal{crit}}$. Each fund $i$ in which $v_i < v_{\textnormal{crit}(i)}$ is added to $Z$.
    
    \item We start the iterative process. At step t, Let $\mathbf{\tilde{b}}_{t-1}$ be a vector with element $\btil_i = \beta_i$ if $i \in Z_{t-1}$ and 0 otherwise, where $\beta_i$ is the loss in value due to failure. Also, for each asset $m_j$ connected with each fund $i$ in $Z_{t-1}$,  its overall market value is reduced as the market's reaction to the fund failure. The price of asset $m_j$ owned by $i$ becomes $p_i = \dfrac{\sum_i w_{ij} - \omega w_{ij} }{\sum_i w_{ij}}$, where $\omega$ is a parameter that measures the strength of the fire sales over the market price of the asset .
    
    \item The new set  $Z_t$ is formed by the funds which  have negative values in: \[ \A \left[\mathbf{W} \vec{1} - \btil_{t-1}  \right] - \mathbf{v}_{\textnormal{crit}}.\]
 
 \item We terminate if $Z_t = Z_{t-1}$. Otherwise, we go back to step 4.
 \end{enumerate}
 
This algorithm provides a framework to understand how damages spread to both other funds and to assets until the cascading failure stops. It also describes a hierarchy of vulnerability under a specific crisis which is determined by the initial shock. 

Many parameters can affect the results of the algorithm. The strength of the initial shock, the assets affected by the shock, the critical value of the funds, the strength of the discontinuous losses and of the fire sales are all inputs of the algorithm and must be determined \textit{ex-ante}.

Next we will present data taken from the Brazilian asset management industry and exemplify the results of the algorithm with simulations.

\section{The Data}

Investment funds , sometimes referred to as collective investment vehicles, are financial intermediaries that collect financial resources from a pool of investors, both individuals and companies, and apply these resources into a pool of assets (\citet{bodie2009investments}).

In Brazil, investment funds are regulated by the Securities and Exchange Comission of Brazil (CVM) through CVM Instruction 555 (ICVM 555) \footnote{ICVM 555 is the current legal diploma  for the regulation of investment funds, but many of the classifications used in this paper are based on the definitions of CVM Instruction 409, which was the legal diploma in the period of the analysis.} \footnote{ICVM 555 regulate both mutual funds and hedge funds, but not Brazilian structured funds (REITs and Receivable funds). Brazilian structured funds are not included in our simulations.}. Investment funds are devoid from legal personality, despite that, they are capable of acquiring and transferring assets and rights, always represented by their administrators and managers (\citet{fortuna2008mercado}). 

Traditionally, funds are classified as fixed income and variable income to discriminate the level of risk of their strategies. Fixed income funds invest in assets with a fixed return rate, such as government bonds and private credit, and variable income funds investing in assets with returns that vary with the market, such as shares of companies. This classification can be refined to better represent investment strategies, two common classifications are provided by the Brazilian Financial and Capital Markets Association (Anbima) and  by CVM. In this work we will adopt the classification used by CVM. 

\begin{figure}[H]
\centering
  \includegraphics[width=0.8\linewidth]{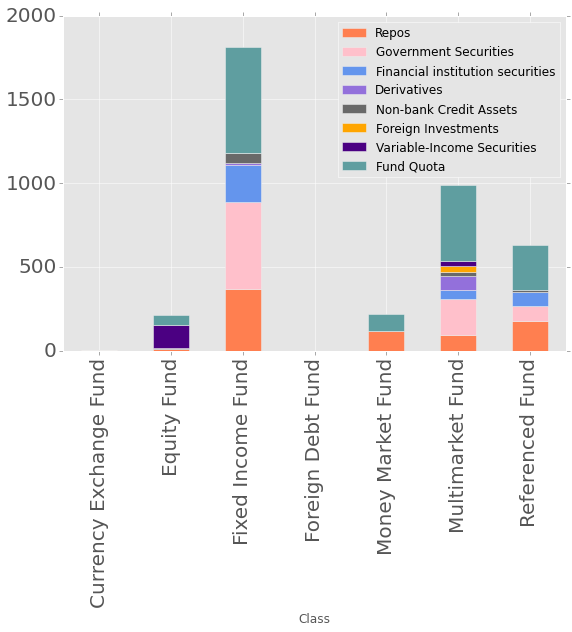}
  \caption{Portfolio composition by fund class. Fixed income funds represent most of the total asset in the industry. Fund quotas is by far the most owned asset, representing a potential source of contagion. Brazilian government bonds follow as second most owned asset.}
  \label{fig:sub1}
\end{figure}

According to CVM's classification, funds are organized in 7 classes which reflect their investment strategies and profile of risk according to the assets they can buy. Fixed-income funds are the most representative class in terms of total assets and Multimarket funds comprise the class with biggest number of funds. Multimarket funds are the class with most diverse strategies and portfolios, as can be seen in figure \ref{fig:sub1}.

Another broadly used classification is for open-ended, which are funds open to redemption at any time after a determined grace period, and closed-ended funds, for funds with strict  restrictions or even completely unavailable for redemption. \footnote{For a full description of classifications of investment funds in Brazil see CVM Instruction 555 and for a broader description of the Brazilian Capital Market see \citet{fortuna2008mercado} and \citet{assaf2001mercado}}. Open-ended funds are the majority both in number of funds and in number of assets under management. Close-ended funds are excluded in simulations, since they are much less susceptible to events that could trigger failure costs, as modeled in this work, such as a run on the fund.

For this study we used data of investment funds from the CVM database. To analyze the network structure we used data from January 2012 up to December 2014 and for the simulations we used data from December 2014.

\begin{figure}[H]
	\centering
  \includegraphics[width=0.8\linewidth]{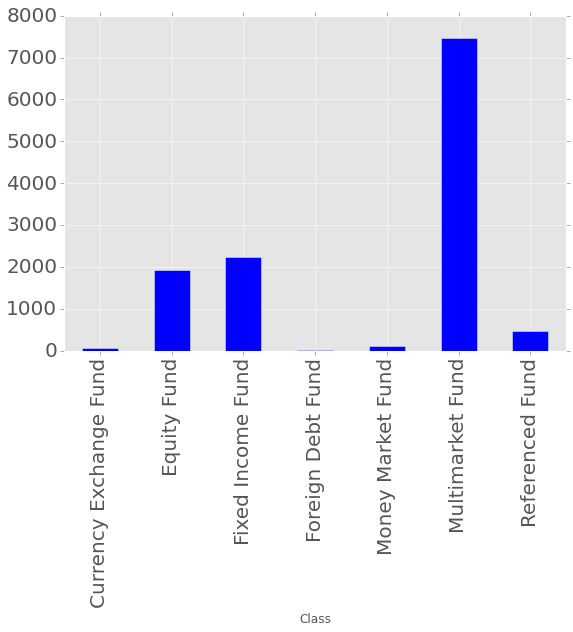}
  \caption{Number of funds by fund class. Multimarket funds represent the largest class in number of funds, followed by fixed-income and equity funds. Funds with more risk appetite are more numerous but smaller in total assets.}
  \label{fig:sub2}
\end{figure}

\section{Results and Discussion}\label{sec:res1}

\subsection{Network Topology}\label{subsec:net1}

To better understand how the market is organized we take a brief look at some key features of the network topology. We then proceed to present the results of the simulation experiments and discuss the role of the topology in the system's dynamics. 

\subsubsection{Stability}

\begin{figure}
	\centering
	\includegraphics[width=0.8\linewidth]{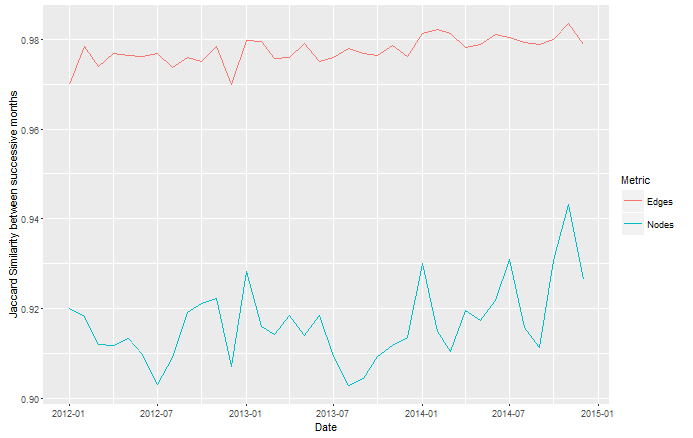}
    \caption{Stability of the Investment Fund network. The table shows Jaccard coeffients of edges and nodes for the network in pairs of months over the period from Jan/2012 to Dec/2014.}
    \label{fig:netstability}

\end{figure}

\begin{figure}
	\centering
    \includegraphics[width=1.0\textwidth]{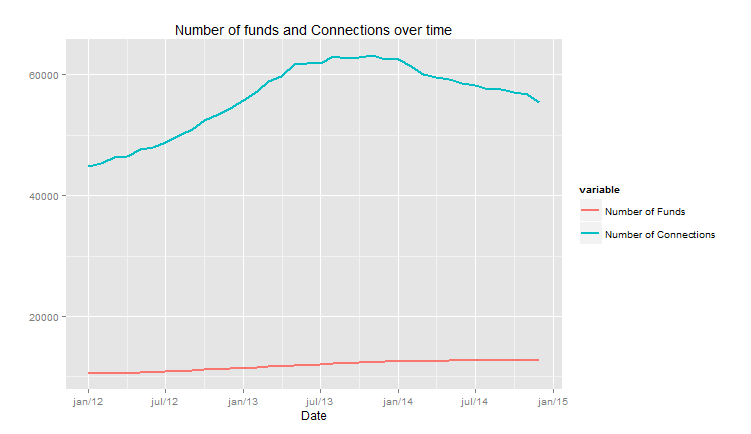}
    \caption{Number of nodes and connections on the network over time. The number over funds grows slowly but steadily over time. The number of edges fluctuates. }
    \label{fig:num funds}
\end{figure}

The structure of the network and it's stability over time reflect investment decisions from asset managers. As in \citet{masys2014networks} Jaccard similarity coefficients are used to measure the structural stability of the network over pairs of successive periods. Figure \ref{fig:num funds} shows the network growth over time and Figure \ref{fig:netstability} reports a summary of the  Jaccard coefficients. The network exhibit considerable stability between successive months, the number of nodes exhibit growth at a steady rate while the number of connections seems to fluctuate more, exhibiting a hump. This fluctuation may indicate some relationship between  the edge count and economic variables which could be further investigated. On average, more then 90\% of connections are stable between periods, which support running   dynamic models (\citet{snijders2010introduction}) such as the cascading failures model.

\subsubsection{Network Metrics}

\begin{figure}[H]
\centering
\begin{subfigure}{.5\textwidth}
  \centering
  \includegraphics[width=.9\linewidth]{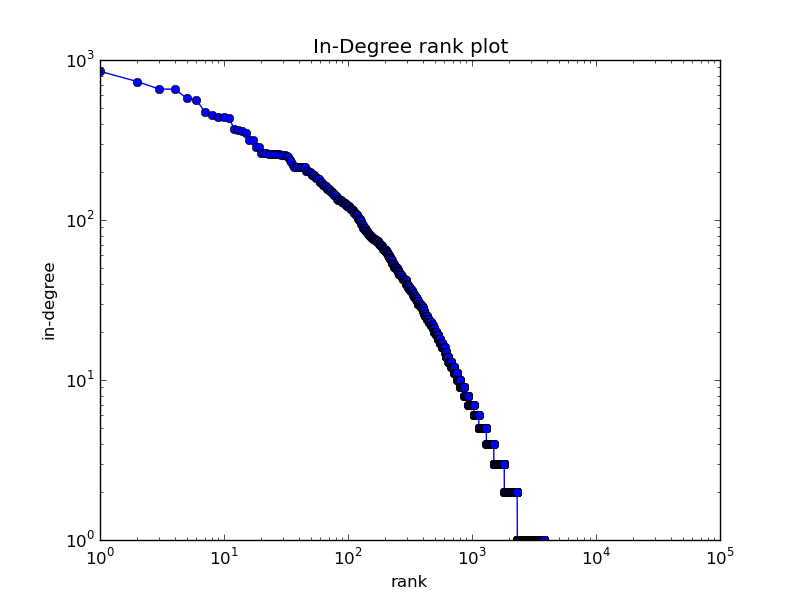}
  \caption{In degree histogram }
  \label{fig:indegree}
\end{subfigure}%
\hfill
\begin{subfigure}{.5\textwidth}
  \centering
  \includegraphics[width=.9\linewidth]{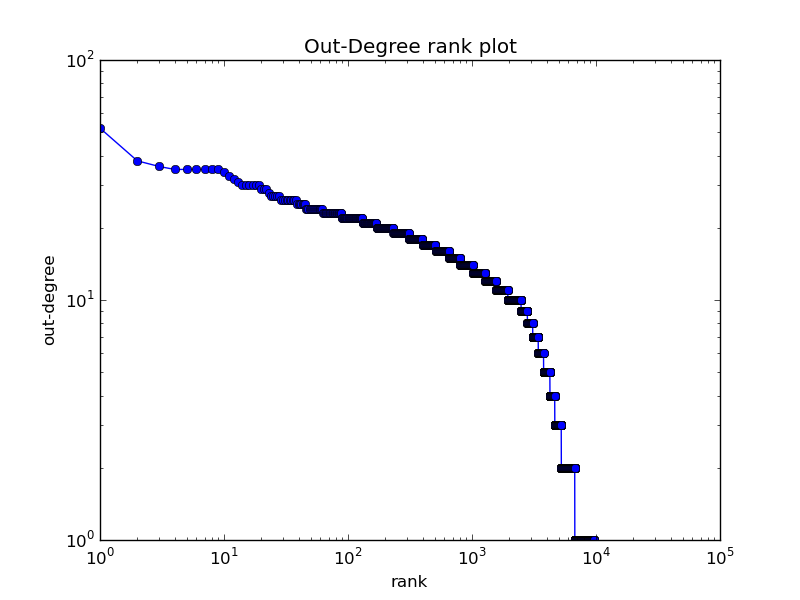}
  \caption{Out degree histogram  }
  \label{fig:outedegree}
\end{subfigure}
\caption{The directionality of the edges indicates the investor fund as the tail and the invested fund as the head. (a) In degree histogram for the cross-holdings network. The histogram exhibit typical power law (b) Out degree histogram for the cross-holdings network. The steepness of the curve is much lower in the out-degree distribution. }
\label{fig:degreehist}
\end{figure}

The cross-holdings network has very low connectivity with an average degree of $d_{\textnormal{avg}} = 4.34$. The degree histogram shows a typical scale free distribution but in-degree and out-degree curves have very different shapes as can be seen in Figure \ref{fig:degreehist}. The highest in-degree is 889 and highest out-degree is 70. The in-degree is the most interesting metric since it is the one that shows how many other funds are directly affected by the spreading of losses in the cross-holdings network.  

\begin{figure}[H]
\centering
\begin{subfigure}{.5\textwidth}
  \centering
  \includegraphics[width=.9\linewidth]{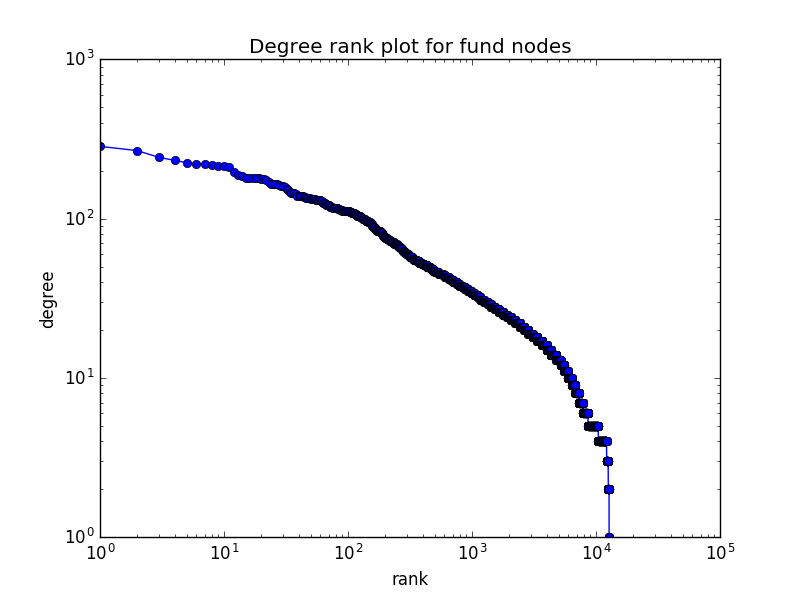}
  \caption{Degree histogram for fund nodes.}
  \label{fig:degfund}
\end{subfigure}%
\hfill
\begin{subfigure}{.5\textwidth}
  \centering
  \includegraphics[width=.9\linewidth]{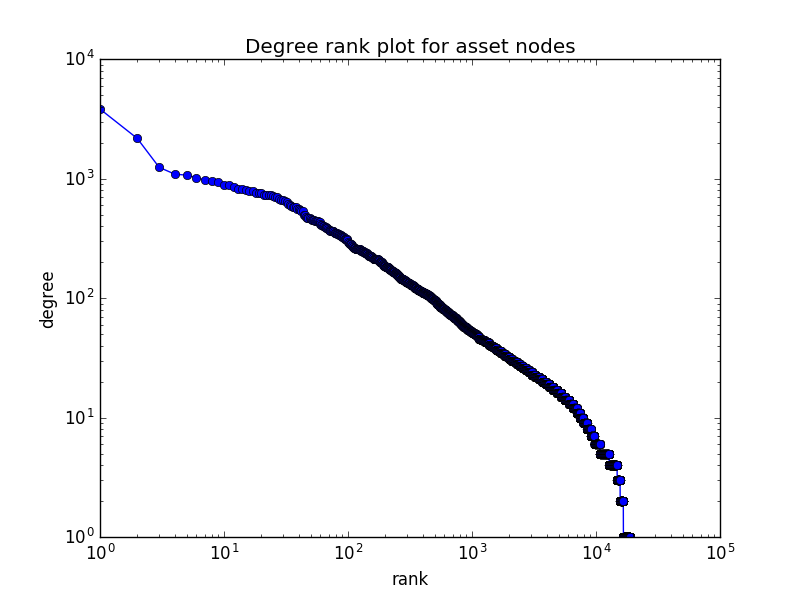}
  \caption{Degree histogram for asset nodes.  }
  \label{fig:degasset}
\end{subfigure}
\caption{Degree histograms for both types of nodes in the bipartite fund-asset network. While both distributions exhibit characteristic power law  shapes, the hubs are much more prominent among assets than funds.  }
\label{fig:bidegreehist}
\end{figure}

The fund-asset network, on the order hand, has an average degree of $d_{\textnormal{avg}} = 20.23$. If we disregard cash, which is connected to almost all funds, the highest asset in-degree of 4838. The degree distribution  in Figure \ref{fig:bidegreehist} shows that a few assets are present in many portfolios displaying a high level of concentration.

%
Comparing the two networks we can observe that the fund-asset network is much more dense, with values of 0.001070 for the fund-asset network and 0.00034 for the cross-holdings network. In Table \ref{tab:centrality} we can see that in the fund-asset network  there are some very central nodes while in the cross-holding network this metric is much weaker. While assortativity is not a relevant metric in the fund-asset network, in the cross-holding network we can observe some level of segregation. Most notably, funds from same Administrators show assortativity of 0.502 and funds from the same class exhibit assortativity of 0.217. While the segregation of funds of the same class is not high enough to indicate regions of confinement for the spread of risk, the segregation among administrators may be an issue of attention.

\begin{table}[ht]
\centering
\caption{Maximum centrality observed in both networks. The Fund-Asset network has nodes with much stronger presence as hubs.}
\label{tab:centrality}
\begin{tabular}{l | c c}
	\hline
	  & Cross-Holdings Network & Fund-Asset Network \\ \hline
	 Max. Degree Centrality & 0.069 & 0.256 \\
	 Max. Closeness Centrality & 0.006 & 0.437 \\
	 Max. Betweenness Centrality & 5.98e-05 & 0.035 \\
 	Max. Eigenvector Centrality & 0.788 & 0.670 \\
	\hline
\end{tabular}
\end{table}
\bigskip

The nature of the spreading process in this model is very different in the cross-holding network and in the bipartite network of funds and assets. The results above indicate that we could observe a faster spreading of contagion caused by asset connections than by cross-holding connections, this is due to the fact that the network is more dense and central assets play a stronger whole as hubs.


\subsection{Simulations}

In this section we illustrate how the use of some network metrics combined with the contagion model can be used to monitor the stability of the financial system and to identify institutions and assets that could rapidly trigger a contagion process. We build our illustration with a financial shock related to the ongoing Brazilian debt crisis, stressing the system to a sovereign debt default.

Sovereign debt default can occur in many forms. A sovereign debt is a contractual obligation and the most clear-cut example of  default is the failure to meet these obligations to pay interest or principal on the due date. Another example is the failure by the government to honor debt it has guaranteed where there are clear provisions for the guarantor to make timely payment.

But sovereign defaults are often not so explicit. Government responses to financial distress can take many forms. In some cases, it can be inferred that, even in the absence of an interruption of debt payments, a default has occurred because actions by the government result in economic losses by creditors,which can vary widely (\citet{beers2014introducing}).

In our network of funds and assets Federal Government Bonds occupy a very central position. It has a total market cap of 35.12\% of the total assets and it also has a degree centrality of 4838,  the biggest among all assets except for Cash. It's average path length 2.19, showing that losses can affect almost any other asset price in the first time step of the algorithm.

In our experimental design we vary the main inputs of the cascading algorithm to test the susceptibility of the network in diverse settings. The most important parameters are the rate of discontinuous loss suffered by funds whenever they fail, the asset prices factor $\omega$, which affects assets of failed funds, the critical value under which the funds will fail and the size of the initial shock.Three results are evaluated: the number of firms failed by the financial shock, the number of total failures caused by the cascading process and the number of iterations before the system reaches a new equilibrium.

\begin{figure}
\centering
\begin{subfigure}{.9\textwidth}
  \centering
  \includegraphics[width=\linewidth]{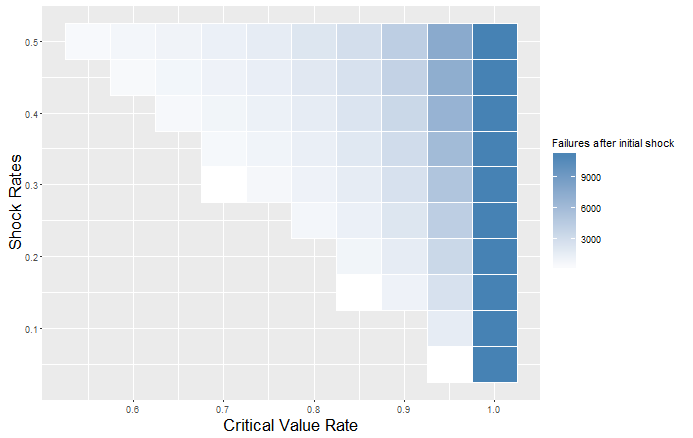}
  \caption{Number of failures after the initial shock. }
  \label{fig:fail_sub1}
 
\end{subfigure}%

\begin{subfigure}{.9\textwidth}
  \centering
  \includegraphics[width=\linewidth]{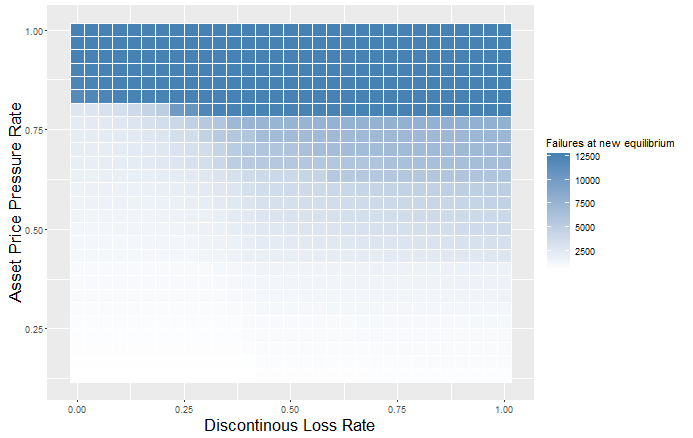}
  \caption{Number of failures at new equilibrium.  }
  \label{fig:fail_sub2}
\end{subfigure}
\caption{(a) Number of failures are we vary both the critical vlaue rate and the shock rate. (b) Number of final failures at a fixed initial shock rate of 15\% and critical value rate of  85\%. }
\label{fig:fail}
\end{figure}

Figure \ref{fig:fail_sub1} shows the number of initial failures caused by the initial shock as we vary the size of the shock and the critical value of organizations. The number of failures is small when the values are close but escalates quickly as the shock becomes much stronger than what investors would tolerate.  Figure \ref{fig:fail_sub2} shows the number of final failures as we vary the discontinuous loss rate and the asset pressure rate  at a fixed initial shock rate of 30\% and critical value rate of  70\%.  The asset pressure rate have a much bigger effect on the number of failures , at a rate of 30\% it leads to a total meltdown of the system independently of the discontinuous loss rate. The discontinuous loss rate does not show the potential to break the whole system on it's own.

These results support our initial hypothesis that the asset network could have a much bigger influence in the final outcome of the contagion due to the network structure and the nature of hubs. We emphasize that we do not see these results as robust, but
merely as illustrative of the dynamic of the process.

\section{Final Remarks}

Financial contagion is a complex phenomena with possibly devastating  consequences to financial systems. Here, extending on previous work from \citet{elliott2014financial} and \citet{huang2013cascading}, we have developed a model that accounts for both cross-holdings  among organizations and pressure over asset prices using two complementary networks. 

The approach we have developed can be a valuable tool for financial supervisors and asset managers. For instance, the algorithm can be used with scenario testing to understand the impact of possible financial crisis to guide supervision and investment decisions. And, while we do not know if the topological assumptions of the model hold for other financial firms, we do believe the framework is still valuable for the study of contagion processes over other financial intermediaries such as banks and insurance companies. 

Although we believe these results are of great value for building a theoretical understanding of financial contagion, the results obtained in this model would hardly be reproducible in a real economy since it doesn't consider interaction with other  financial intermediaries, existing regulatory measures and  the direct intervention of financial regulators and/or government bailout.

Several improvements to the modeling process are possible to make it more realistic and closer to the observable reality. To advance this line of research: (1) More work is required to understand critical values under which discontinuities occur and the value of the loss caused by these discontinuities, (2) the effect of bankruptcy over asset prices needs to be better determined, (3)  a better model of financial shocks should be explored, (4) the interaction with other financial intermediaries should be included and, (5) The effects of regulation should be considered.


\bibliographystyle{apalike}
\bibliography{sample}


\end{document}